\documentclass[aps,prd,tightenlines,groupedaddress,showpacs,
preprintnumbers,nofootinbib,nobibnotes,notitlepage,amsmath,amssymb, 
onecolumn, 11pt, letterpaper, floatfix]{revtex4-1}

\usepackage[utf8]{inputenc}
\usepackage[normalem]{ulem}
\usepackage{graphicx}
\usepackage{dcolumn}
\usepackage[colorlinks=true,allcolors=purple]{hyperref}
\usepackage{url}
\usepackage{enumerate}
\usepackage{comment}

\usepackage{slashed,multirow,relsize,soul,feynmp-auto,tikz}
\usepackage{color}
\usepackage{mathrsfs} 
\usepackage{amsmath}
\usepackage{cancel}
\usepackage{bbold}
\usepackage{mathrsfs}
\usepackage{braket}
\usepackage{physics}
\usepackage{siunitx}
\usepackage{multirow}
\usepackage[capitalize]{cleveref}
\usepackage{xspace}
\usepackage{float}

\usepackage{slashed,multirow,relsize,soul,feynmp-auto,tikz}
\usepackage{color}
\usepackage{mathrsfs} 
\usepackage{amsmath}
\usepackage{cancel}
\usepackage{bbold}
\usepackage{mathrsfs}
\usepackage{braket}
\usepackage{physics}
\usepackage{siunitx}
\usepackage{multirow}
\usepackage[capitalize]{cleveref}
\usepackage{xspace}
\usepackage{float}

\hypersetup{
  colorlinks  = true,
  urlcolor    = blue,
  linkcolor   = blue,
  citecolor   = red
}

\definecolor{kjkblue}{rgb}{0.39, 0.589, 0.6914}

\newcommand{\efours}{$\left\lvert U_{e4}\right\rvert^2$}
\newcommand{\mufours}{$\left\lvert U_{\mu 4}\right\rvert^2$}
\newcommand{\taufours}{$\left\lvert U_{\tau 4}\right\rvert^2$}
\newcommand{\orcid}[1]
{\begingroup
  \hypersetup{hidelinks}\href{https://orcid.org/#1}{\includegraphics[width=9pt]{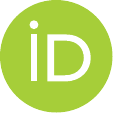}
} \endgroup}

\begin{document}

\preprint{MI-HET-884}

\title{Sterile Neutrino Mixing Parameters from Solar-Neutrino Coherent Scattering}

\author{Kevin J. Kelly \orcid{0000-0002-4892-2093}}
\email{kjkelly@tamu.edu}
\affiliation{Department of Physics and Astronomy, Mitchell Institute for Fundamental Physics and Astronomy, Texas A\&M University, College Station, Texas 77843, USA}
\author{Nityasa Mishra \orcid{0009-0005-1519-8093}}
\email{nityasa\_mishra@tamu.edu}%
\affiliation{Department of Physics and Astronomy, Mitchell Institute for Fundamental Physics and Astronomy, Texas A\&M University, College Station, Texas 77843, USA}
\author{Louis E. Strigari \orcid{0000-0001-5672-6079}}%
\email{strigari@tamu.edu}%
\affiliation{Department of Physics and Astronomy, Mitchell Institute for Fundamental Physics and Astronomy, Texas A\&M University, College Station, Texas 77843, USA}
\date{\today}

\begin{abstract}
Recently, dark matter direct-detection experiments have begun their exploration of the ``neutrino fog,'' providing the first hints of detection of solar neutrinos scattering elastically with the nuclei in the detector.
In this work, we investigate how such observations can be used to uniquely explore sterile-neutrino parameter space, specifically through mixing with $\nu_\mu$ and $\nu_\tau$.
While it is challenging to constrain these parameters with current observations -- PandaX, XENONnT, and LZ -- we demonstrate how future measurements (with modest improvements to exposure and systematic uncertainties) can provide useful, complementary information in the search for sterile neutrinos.
With an ideal, next-generation direct-detection facility (${\sim}3000$ ton-yr), we can probe parameter space previously unexplored by other methods, including long-baseline and atmospheric searches for this class of new physics.
\end{abstract}

\maketitle

\vspace{-2em}
\section{Introduction \label{sec:introduction}}
Experiments designed to search for dark matter through direct-detection processes, in which dark matter scatters with electrons and/or nuclei in low-threshold detectors, continue to improve their sensitivity to unprecedented levels~\cite{PandaX:2024qfu,XENON:2024hup,LZ:2024zvo,XENON:2026ydt}.
The experimental sensitivity is now improving to the point at which they are able to detect to Solar neutrinos, reaching the so-called ``neutrino fog''~\cite{Billard:2013qya,Ruppin:2014bra,OHare:2021utq}.
As experimental exposures increase, these measurements may be used to better understand neutrino sources, and possibly neutrino properties themselves~\cite{Dutta:2019oaj,Mishra:2023jlq}.

Neutrinos mimic dark-matter-scattering signals in these environments when they scatter coherently off the nucleus via the CE$\nu$NS (coherent elastic neutrino-nucleus scattering) process~\cite{Abdullah:2022zue}.
At tree level, this process is equally sensitive to scattering by all active neutrino flavors through exchange of the $Z$ boson, which higher order corrections providing some flavor dependence~\cite{Sehgal:1985iu,Tomalak:2020zfh}.
The first detection of neutrinos by dark-matter detectors was recently reported by the PandaX-4T~\cite{PandaX:2024muv}, XENONnT~\cite{XENON:2024ijk}, and LZ~\cite{LZ:2025igz} collaborations, each identifying $^8$B solar neutrinos as the source.
These experiments operate with effective nuclear recoil energy thresholds $\lesssim 1$ keV, and given their exposures and backgrounds, they identify the $^8$B component of the solar neutrino flux at a the $\sim 2\sigma - 4\sigma$ significance level.
This is the first detection of CE$\nu$NS from an astrophysical source, complementing the detections from the stopped-pion source by the COHERENT experiment~\cite{COHERENT:2020iec,COHERENT:2021xmm,COHERENT:2024axu}, and the CONUS+ detection from a nuclear reactor source~\cite{Ackermann:2025obx}.

Previous studies have considered the prospects for studying beyond the standard model (BSM) physics using PandaX-4T, XENONnT, and LZ~\cite{AristizabalSierra:2024nwf,Blanco-Mas:2024ale,Li:2024iij,DeRomeri:2024iaw,Duque:2025mno,DeRomeri:2026prc}.
These analyses showed that, since all three neutrino flavors are contained within the $^{8}$B flux, these experiments are uniquely sensitive to non-standard neutrino interactions in each of the electron, muon, and tau neutrino sectors.
The measurements additionally provide new insights into standard model (SM) neutrino physics~\cite{DeRomeri:2024hvc}. 
In this work, we will focus on a specific BSM scenario in the context of measuring solar neutrinos using dark matter detectors and the CE$\nu$NS process: the scenario in which a new, light sterile neutrino exists.

CE$\nu$NS is sensitive to all active neutrino flavors interacting; in the context of a sterile neutrino, we are sensitive to the possibility where electron neutrinos, produced in the Sun's core, oscillate into the sterile eigenstate before interacting in the dark matter detector. 
Even though this detection method cannot identify $\nu_\mu$ and $\nu_\tau$ (e.g., via charged-current interactions), these oscillation probabilities can still be sensitive to the mixing between $\nu_{\mu,\tau}$ and the new mass eigenstate~\cite{Dentler:2018sju}.
Because of this unique sensitivity, as well as the fact that $\nu_e$-sterile mixing is well constrained, we will focus on the mixing parameters $|U_{\mu 4}|^2$ and $|U_{\tau 4}|^2$ throughout this work.
Earlier studies have examined the prospects for identifying sterile neutrinos at dark matter detectors in the simple case of one free mixing parameter~\cite{Billard:2014yka}.
Despite strong constraints from other search strategies~\cite{Dentler:2018sju}, we will demonstrate how CE$\nu$NS detection of solar neutrinos in dark matter detectors can be highly complementary.
While current PandaX-4T, XENONnT, and LZ results cannot improve on current constraints, we find that with reasonable increases in experimental exposure and improvements to systematic uncertainties, direct detection facilities can probe previously untested sterile neutrino parameter space.

The remainder of this work is organized as follows:~\cref{sec:Sterile} introduces the sterile neutrino framework on which we focus here and its impact on the solar neutrino flux present at earth. In~\cref{sec:CEvNS}, we briefly discuss the CE$\nu$NS scattering process and how we reinterpret the existing experimental results. We spend some time focusing on current systematic uncertainties as a means of understanding future capability, which we expound upon in ~\cref{sec:Future}. In~\cref{sec:Conclusions}, we offer some concluding remarks.

\section{Sterile Neutrino Impact on Solar Neutrino Flux}
\label{sec:Sterile}
We are interested in the scenario where a light ($\lesssim$ keV) sterile neutrino exists that mixes with the mostly-active neutrinos -- in this case, we extend the $3\times3$ mixing matrix to be $4\times4$, and three new physical mixing angles are relevant.
We will work in terms of the mixing-matrix elements $\left\lvert U_{\alpha 4}\right\rvert^2$, $\alpha = e,\, \mu,\, \tau$.\footnote{Two additional CP-violating phases are present in this case, however we are not sensitive to their effect and so we disregard them for this analysis.}
At the sun's core, only $\nu_e$ are produced; our goal is to determine the flavor structure of those neutrinos as they exit the sun and reach the earth.

Solar neutrinos have energies well below the charged muon and tau masses, and so the only detection methods are through either neutral-current or charged-current $\nu_e$ scattering.
To that end, we require only two effective probabilities for our calculation: $P(\nu_e \to \nu_e)$, i.e., the survival probability of electron neutrinos when they arrive at a detector on earth, and $P(\nu_e \to \nu_s)$, i.e., the fraction of electron neutrinos that are no longer `active' when reaching the detector. 
We can express these effective probabilities using
\begin{align}
    P\left(\nu_e \to \nu_z\right) = \sum_{k=1}^{4} \left\lvert U_{ek}^{m}\right\rvert^2 \left\lvert U_{zk}\right\rvert^2,
\end{align}
where $z = e$ or $s$ (the `sterile' flavor combination), and the superscript `$m$' indicates the matter-induced values of the mixing-matrix elements for the neutrinos as they propagate from the core of the sun to its surface.
The elements $\left\lvert U_{sk}\right\rvert^2$, which are unphysical, may be determined by imposing the unitarity of the $4\times4$ mixing matrix.
We will work in the regime where the sterile neutrino is sufficiently massive so that the new mass eigenstate ``decouples'' from the others and evolves adiabatically. This requires approximately $\Delta m_{41}^2 \gtrsim 10^{-4}$~eV$^2$, which we will assume throughout the remainder of this work.

In this approximation, we can derive analytic expressions for the two effective probabilities, which are useful in understanding the results to come.
\begin{align}
    P\left(\nu_e \to \nu_e\right) &= \left( c_{13}^4 P_{ee}^{\rm SM} + s_{13}^4\right)\left( 1 - \left\lvert U_{e4}\right\rvert^2\right) + \left\lvert U_{e4}\right\rvert^2, \label{eq:Pee_Approx}\\
    P\left(\nu_e \to \nu_s\right) &= \left(\left(c_{12}^{m}c_{13}\right)^2 \left\lvert U_{s1}\right\rvert^2 + \left(s_{12}^m c_{13}\right)^2 \left\lvert U_{s2}\right\vert^2 + s_{13}^2 \left\lvert U_{s3}\right\rvert^2 \right) \left(1 - \left\lvert U_{e4}\right\rvert^2\right) + \left\lvert U_{e4}\right\rvert^2 \left \lvert U_{s4}\right\rvert^2     \label{eq:Pes_Approx}
\end{align}
Here, $P_{ee}^{\rm SM}$ is the expected solar-$\nu_e$ survival probability in the absence of a new sterile neutrino eigenstate;~\cref{eq:Pee_Approx} provides us a way of interpreting deviations from this expectation driven by the new state.
We have explicitly substituted the electron-row mixing elements in terms of their standard mixing-angle representation, $c_{ij} = \cos\theta_{ij}$, $s_{ij} = \sin\theta_{ij}$, with $\theta_{12}$ and $\theta_{13}$ being two of the three mixing angles present in three-neutrino oscillations.
For this approximation, we make the simplifying assumption that $\theta_{12}$ is the only mixing angle modified by the matter effects in the sun.

A great deal of work has previously explored the use of solar neutrinos to constrain a sterile neutrino mixing with $\nu_e$ via nonzero \efours -- by contrasting~\cref{eq:Pee_Approx} and~\cref{eq:Pes_Approx}, we see that $P(\nu_e \to \nu_e)$ may readily probe nonzero \efours -- this can be done stringently using, e.g., SNO, Super-Kamiokande, and Borexino's results for $\nu_e$ charged-current scattering. We refer the interested reader to Ref.~\cite{Gonzalez-Garcia:2024hmf,Goldhagen_2022} and references therein for thorough discussion of $\nu_e$-based constraints on \efours in the context of a light sterile neutrino.

\begin{figure}
    \centering
\includegraphics[width=0.49\linewidth]{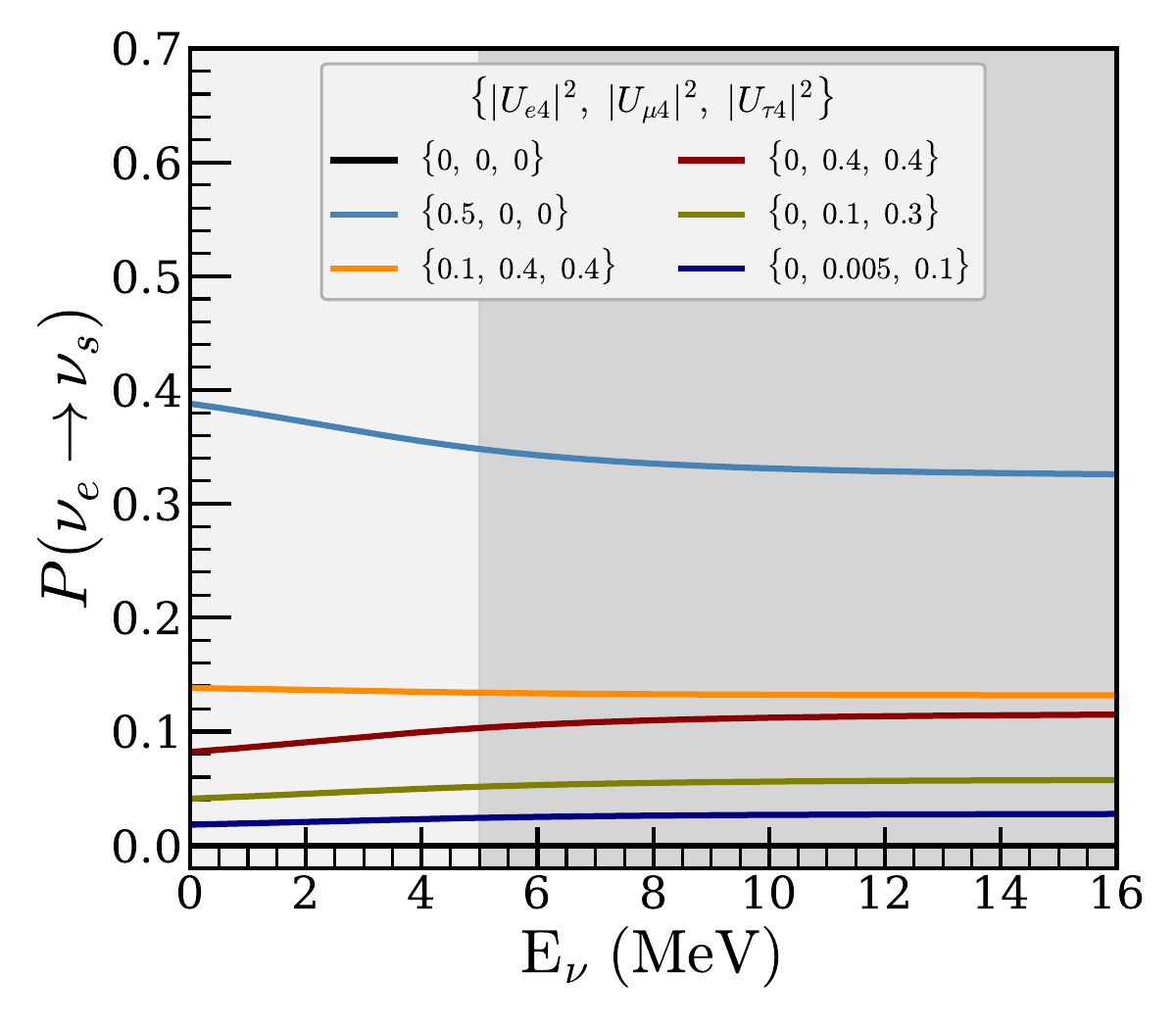}
\includegraphics[width=0.49\linewidth]{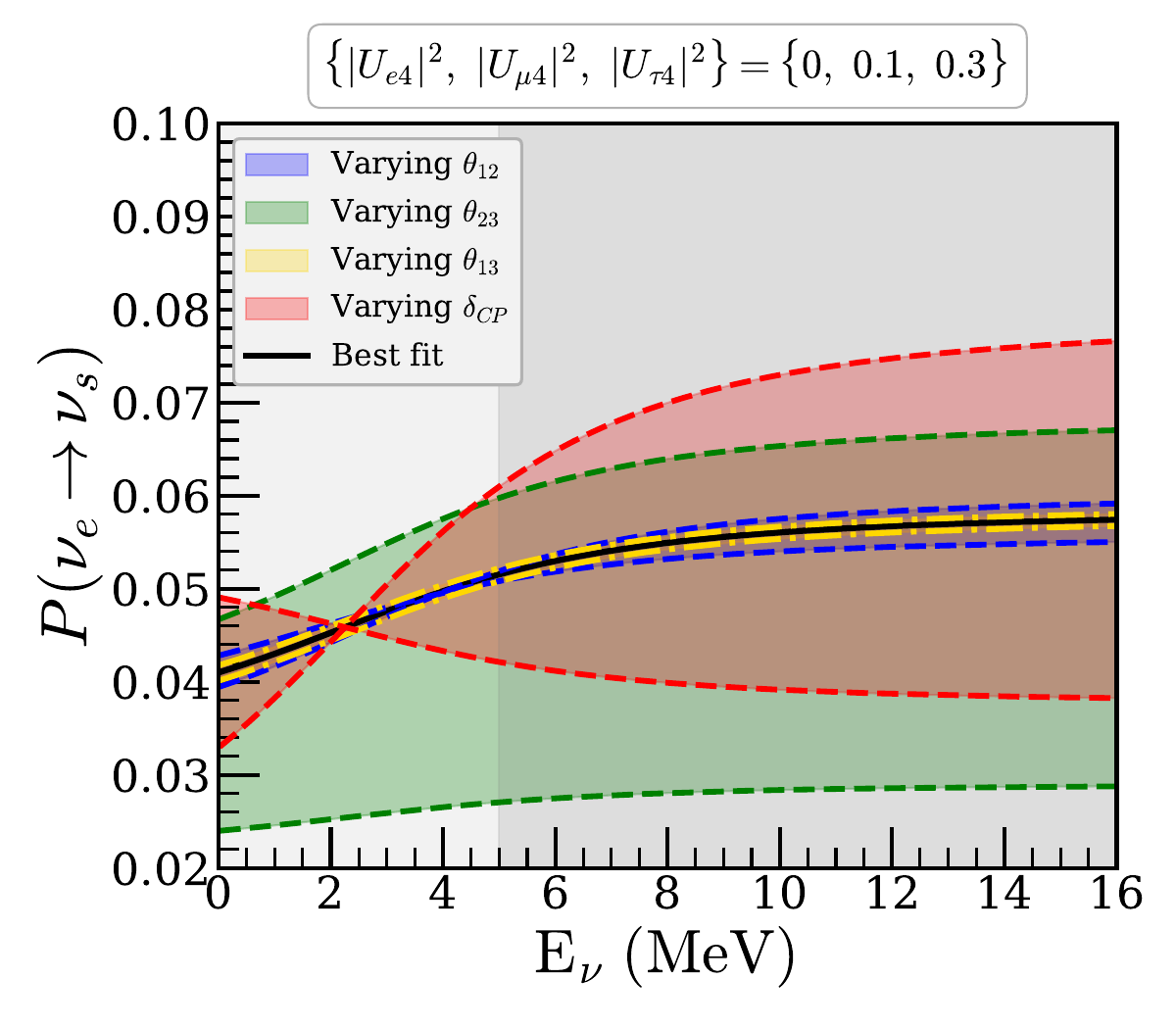}
    \caption{\textit{Left:} Oscillation probability of electron neutrinos produced in the sun into sterile neutrinos at detectors on earth as a function of neutrino energy for a variety of choices of new-physics mixing parameters. The gray, shaded region indicates the energy range of interest for solar neutrinos interacting in direct-detection experiments.
    \textit{Right:} For a fixed combination of new-physics parameters ($\left\lvert U_{e4}\right\rvert^2 = 0$, $\left\lvert U_{\mu 4}\right\rvert^2 = 0.1$, $\left\lvert U_{\tau 4}\right\rvert^2 = 0.3$), the impact of varying different ``standard'' mixing parameters on the oscillation probability $P(\nu_e \to \nu_s)$. Each parameter depicted is allowed to vary, one-at-a-time, within its current $3\sigma$ allowed range~\cite{Esteban:2024eli}.\label{fig:probability_Pee_and_Pes}}
\end{figure}

Instead, we turn our attention towards the capabilities of CE$\nu$NS for probing a light sterile neutrino in a direct-detection experiment. In the limit $\left\lvert U_{e4}\right\rvert^2 \to 0$, the oscillation probability $P(\nu_e \to \nu_e)$ approaches its SM expectation and $\nu_e$ probes are completely insensitive to the potential existence of a sterile neutrino. In this scenario, $P(\nu_e \to \nu_s)$ probes are the only mechanism to test \mufours and \taufours.
When considering neutral-current scattering via the coherent-scattering CE$\nu$NS process, we will predominantly be sensitive to incoming neutrinos with energies $5$~MeV $\lesssim E_{\nu} \lesssim 16$~MeV. In~\cref{fig:probability_Pee_and_Pes} (left) we demonstrate the oscillation probability $P(\nu_e \to \nu_s)$ across this energy range for a handful of parameter combinations. We find that, for all parameter space of interest, this effective oscillation probability is flat, which will be a key factor (even a hindrance) in the analysis to follow.

\begin{figure}
    \centering
\includegraphics[width=0.49\linewidth]{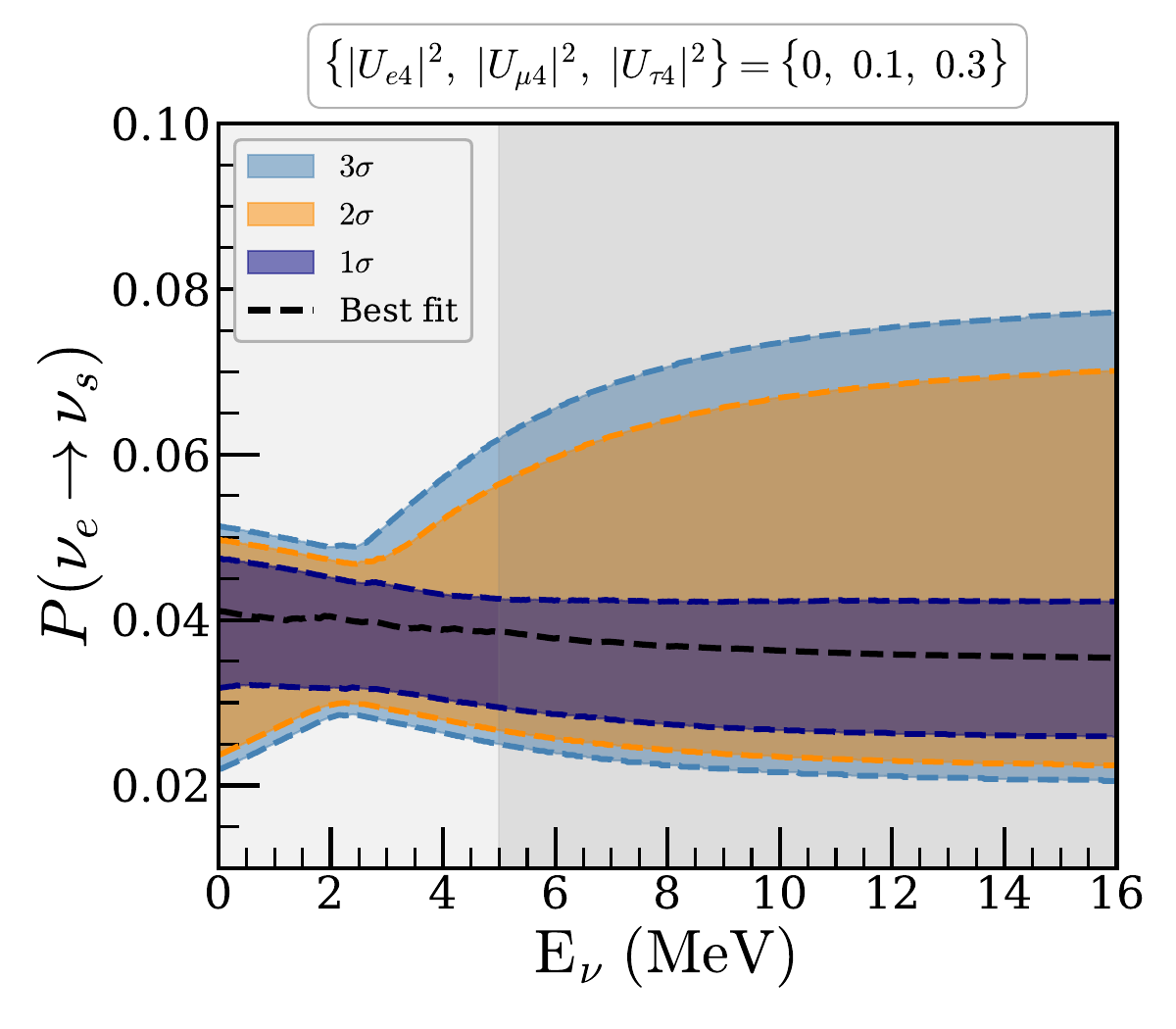}
\includegraphics[width=0.49\linewidth]{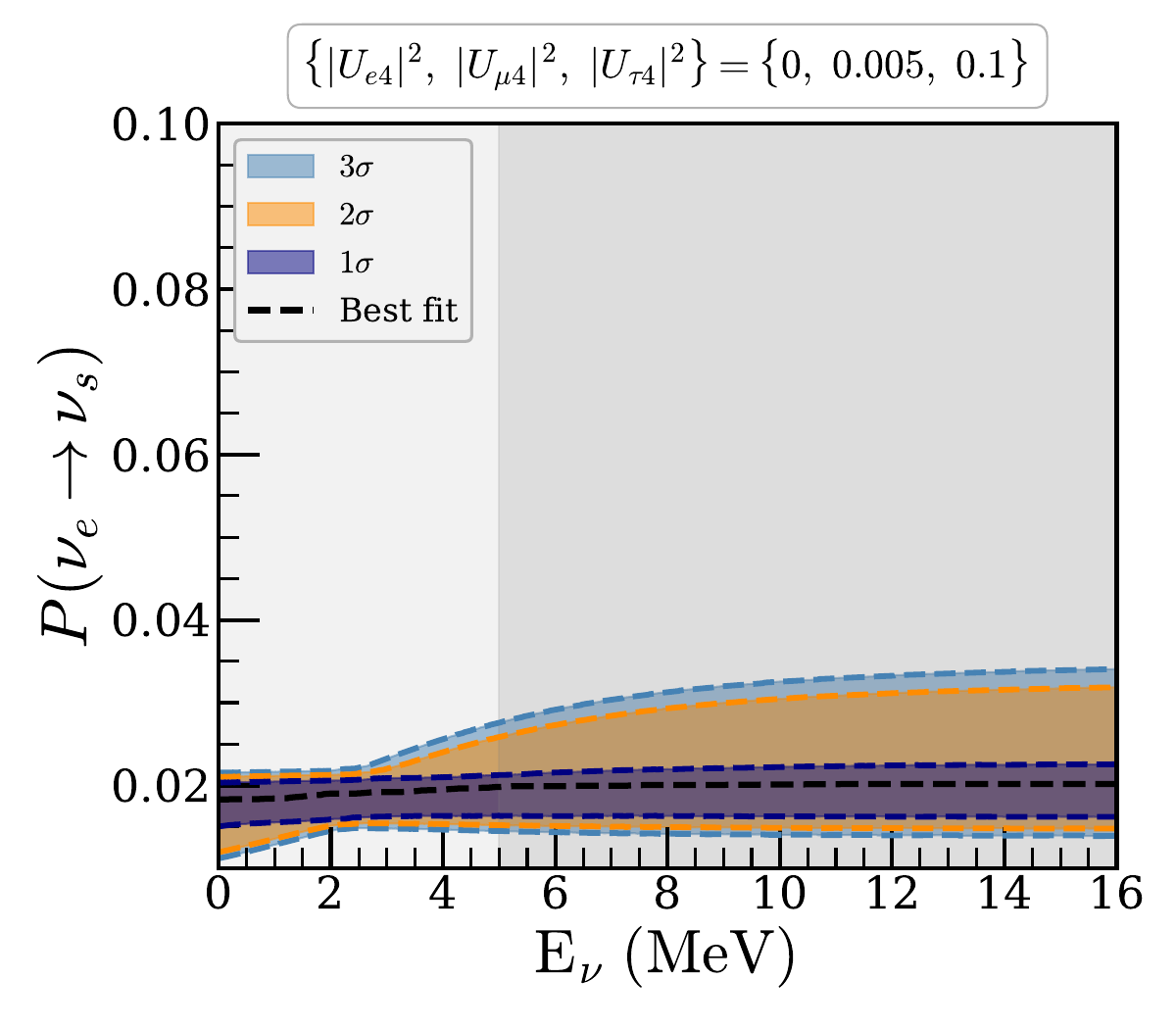}
    \caption{The impact of varying standard, three-flavor oscillation parameters on the oscillation probability $P(\nu_e \to \nu_s)$ for new-physics parameters corresponding to Point A (left) and Point B (right), as labelled above each panel. The different color bands correspond to the $1$, $2$, and $3\sigma$ CL expectation of the oscillation probabilities if we allow all oscillation parameters to vary according to \textit{current} knowledge~\cite{Esteban:2024eli}.\label{fig:Pes_UncertaintyImpact}}
\end{figure}
While we will attempt to constrain the new mixing parameters \mufours and \taufours, we note that their impact on $P(\nu_e\to\nu_s)$ depends significantly on the ``standard'' oscillation parameters entering~\cref{eq:Pes_Approx}. The right panel of~\cref{fig:probability_Pee_and_Pes} demonstrates this explicitly, where (fixing \efours$= 0$, \mufours$=0.1$, \taufours$=0.3$) we allow four of the standard mixing parameters to vary within their $3\sigma$ currently-allowed range, per~\cite{Esteban:2024eli}. 
Since $\theta_{12}$ and $\theta_{13}$ are already well constrained, varying them makes little impact on $P(\nu_e \to \nu_s)$ -- the largest impact comes from uncertainty on $\theta_{23}$ (given that it is the largest mixing angle and is less precisely measured) and $\delta_{\rm CP}$.

Through the remainder of this work, we will be interested in two specific new-physics benchmark points,
\begin{align}
    \mathrm{Point\ A}:\quad \left\lbrace \left\lvert U_{e4}\right\rvert^2,\ \left\lvert U_{\mu 4}\right\rvert^2,\ \left\lvert U_{\tau 4}\right\rvert^2\right\rbrace &= \left\lbrace 0,\ 0.1,\ 0.3\right\rbrace, \label{eq:PointA} \\
    \mathrm{Point\ B}:\quad \left\lbrace \left\lvert U_{e4}\right\rvert^2,\ \left\lvert U_{\mu 4}\right\rvert^2,\ \left\lvert U_{\tau 4}\right\rvert^2\right\rbrace &= \left\lbrace 0,\ 0.005,\ 0.1\right\rbrace. \label{eq:PointB}
\end{align}
Point A is representative of parameter space that is currently testable by other means of studying solar neutrinos; Point B is currently untested by any experimental approach~\cite{Dentler:2018sju}.
For nominal values of the standard oscillation parameters, we expect that $P(\nu_e \to \nu_s) \approx 0.04-0.05$ ($0.02-0.03$) for Point A (B).
If we extend the analysis of~\cref{fig:probability_Pee_and_Pes} (right) to allow all oscillation parameters to vary within their current uncertainties~\cite{Esteban:2024eli},\footnote{Specifically, we use the results of Ref.~\cite{Esteban:2024eli} to draw random sets of standard oscillation parameters, following the covariances in parameters evident in the $\chi^2$ tables provided. This leads to some discrepancy between~\cref{fig:Pes_UncertaintyImpact}(left) and~\cref{fig:probability_Pee_and_Pes}(right), in which we assumed $\delta_{\rm CP} = -\pi/2$.}, we find that even at $2\sigma$~CL, these probabilities can vary by a factor of two.
This is demonstrated explicitly in~\cref{fig:Pes_UncertaintyImpact} for Point A (left) and Point B (right).
As with~\cref{fig:probability_Pee_and_Pes} (right), most of this variance comes from the uncertainty on $\theta_{23}$ and $\delta_{\rm CP}$; with future precision measurements of these quantities from DUNE and T2HK, the size of the colored bands in~\cref{fig:Pes_UncertaintyImpact} should decrease substantially.
For the remainder of this work, we will neglect the impact of standard-parameter uncertainties on the new-physics sensitivity.
In principle, this could be incorporated as a systematic uncertainty that weakens our projected sensitivity.

\section{Coherent Scattering of Solar Neutrinos }
\label{sec:CEvNS}
Dark matter direct detection experiments can observe solar neutrinos via flavor-blind, neutral-current Coherent Elastic neutrino Nucleus Scattering (CE$\nu$NS) process. As a result, these experiments are equally (up to radiative corrections~\cite{Tomalak:2020zfh}) sensitive to all active neutrino flavors. In the context of a sterile neutrino analysis, this means they are capable of measuring the quantity $\left[1 - P(\nu_e\rightarrow\nu_s)\right]$~\eqref{fig:probability_Pee_and_Pes}.

The differential cross-section of this coherently enhanced scattering process is given by
\begin{equation}
    \frac{d \sigma}{dT_N} = \frac{G_F^2 M_N}{\pi}Q^2_W \left[1 - \frac{T_N}{E_\nu} - \frac{M_N T_N}{2 E_\nu^2} \right] F^2_W(Q^2), 
\end{equation}
where $E_\nu, T_N$ and $M_N$ signify neutrino energy, nuclear recoil energy, and mass of the target nucleus, respectively. $G_F$ is the Fermi constant and $Q_W = \frac{1}{2}\left(\left(1 - 4 \sin^2{\theta_W}\right)Z - N\right)$ is the weak charge that depends on the number of protons ($Z$), the number of neutrons ($N$) and the sine of the weak mixing angle $\theta_W$. $F_W(Q^2)$ is the nuclear form factor dependent on the momentum transferred to the nucleus $Q$.

For this analysis, we focus our attention on current experiments that have observed $^8$B solar neutrinos with greater than $2\sigma$ significance, namely PandaX-4T, XENONnT, and LZ.
Each is a Xenon detector utilizing a dual-phase time-projection chamber (TPC), producing both scintillation photons (S1 signal) associated with nuclear recoils and corresponding ionization electrons (S2 signal).
A summary of key experimental details relevant for this work is shown in~\cref{tab:table1}.
\begin{table}[h!]
  \begin{center}
\renewcommand{\arraystretch}{1.3}
    \begin{tabular}
    {|l|c|c|c|c|} 
    \hline
       & \textbf{PandaX-4T Paired} & \textbf{PandaX-4T US2} & \textbf{XENONnT Paired} & \textbf{LZ Paired} \\
      \hline
      Threshold & 1.1 keV & 0.33 keV & 0.5 keV & 1.0 keV \\
      \hline
    Exposure & 1.20 (t$\times$y) & 1.04 (t$\times$y)  & 3.51 (t$\times$y) & 5.7 (t$\times$y) \\
    \hline
    \end{tabular}
    \caption{Experimental thresholds and exposures for different data sets.}
     \label{tab:table1}
  \end{center}
\end{table}

In Ref.~\cite{PandaX:2024muv}, two data sets have been analyzed by the PandaX-4T Collaboration; a combined S1/S2 analysis (``Paired'') or a S2-only one (``US2'').
The reported number of events observed (including backgrounds) for the two data sets is 3 and 332 respectively, corresponding to joint likelihood best-fit values for $^8$B signal of 3.5$\pm$1.3 and 75$\pm$28 with fully correlated uncertainties.
In addition, they also present the unblinded US2 data set plotted against the number of ionized electrons, providing spectral information of two dominant sources of background: the radioactivity on the cathode electrode referred to as cathode background and ``Micro-discharging" (MD). For analyzing the spectral data we adopt the procedure presented in Ref~\cite{DeRomeri:2024iaw}.

For the case of XENONnT~\cite{XENON:2024ijk}, the experiment observed 37 observed paired events corresponding to a best-fit value of $10.7^{+3.7}_{-4.2}$ $^8$B events. The dominant backgrounds considered for the data set are accidental coincidence (AC), radiogenic neutrons originating from the detector (neutron background) and beta decays from radioactive impurities and electrons scattered by external photons (ER background). 

Finally, LZ~\cite{LZ:2025igz} also studies paired-coincidence events with both S1 and S2 energy depositions. Accounting for accidental coincidence backgrounds as well as neutron-induced events, the total expected background rate is approximately $6.64$~events, where a total of $19$~were observed. The expected rate of $^{8}$B CE$\nu$NS events was $20.6$, leading to a total expected rate of $27.2$. The collaboration performed a fit, normalizing the $^{8}$B flux in their analysis, and obtained a value below the expectation -- $12.3^{+7.0}_{-5.4}$ events. 

Putting the three experiments in context, the size of the uncertainties on the measured solar neutrino flux using CE$\nu$NS is roughly comparable, with LZ reporting the most precise measurement (despite the smaller-than-expected observation). There is mild ($\lesssim 1\sigma$) disagreement between the LZ and PandaX-4T results, with XENONnT's measurement being compatible with each of the other two.

We showed in~\cref{fig:probability_Pee_and_Pes} that the $^{8}$B flux is only sensitive to overall normalization effects in the oscillation probability $P(\nu_e \to \nu_s)$ in the context of a light sterile neutrino. Neverthelesss, we may calculate the expected event rate in a given energy bin using
\begin{equation}
    \text{R}_i = \zeta^{\rm exp.} C^{\rm exp.}_i \int_i \left[ \int \left[1 - P(\nu_e\rightarrow\nu_s)\right]\frac{d \phi}{d E_\nu}\frac{d\sigma}{d T_N} dE_\nu\right] \varepsilon^{\rm exp.}\frac{d T_N}{d n^{\rm exp.}} dn^{\rm exp.}\, ,
    \label{eq: event rate}
\end{equation}
where superscript ${\rm exp.}$ refers to a particular experiment and the subscript ${i}$ signifies the $i^{th}$ bin. $\zeta^{\rm exp.}$ is the experiment-dependent exposure given in~\cref{tab:table1}, $\varepsilon^{\rm exp.}$ is the signal efficiency/ signal acceptance for the experiment extracted from Refs.~\cite{XENON:2024ijk,PandaX:2024muv,LZ:2025igz} and $\frac{d \phi}{d E_\nu}$ gives the solar neutrino flux information. In addition to these, we also include correction factors $C^{\rm exp.}_i$ to match our standard model predictions (when $P(\nu_e\rightarrow\nu_s) = 0 $) with the best-fit results presented by the collaborations. This is done in order to account for effective efficiencies and detector effects. 

The observed data for XENONnT are binned in terms of 3 bins of ``cS2" (corrected S2 photoelectrons) in the range [120, 500], while PandaX-4T binned the data in terms of number of S2 electrons $N_{e^-}$ with signal data dominating 8 bins in the range [4, 8]. As a result, the event rate have to be calculated in terms of $n^{X,P}$ (in~\cref{eq: event rate}) which are functions of $T_N$ (see~\cref{tab:table2}). 
\begin{table}[h!]
  \begin{center}
\renewcommand{\arraystretch}{1.3}
    \begin{tabular}
    {|c|c|c|} 
    \hline
       & \textbf{PandaX-4T US2 only} & \textbf{XENONnT Paired}\\
      \hline
      number of bins & 8 & 3  \\
      \hline
   $n^{X,P}$ & Number of S2 $e^-$ & cS2 PE \\
    \hline
    range $n^{X,P}$ & [4,8] & [120,500] \\
    \hline 
    $n^{X,P}(T_N)$ & $T_N Q^P_y(T_N)$ & $T_N Q^X_y(T_N)g2$ \\
    \hline
    \end{tabular}
    \caption{$Q^{X,P}_y(T_N)$ is the charge yield extracted from Refs.~\cite{XENON:2024xgd,PandaX:2024muv} and g2 = 16.9 PE/electron is the electron gain constant.}
     \label{tab:table2}
  \end{center}
\end{table}
\vspace{-3em}

\subsection{Systematic Uncertainties in Current Analyses}
In order to address sensitivity to the signal of a sterile neutrino which modifies $P(\nu_e \to \nu_s)$, we need to account for systematic uncertainties in these direct-detection-as-CE$\nu$NS facilities.
Broadly speaking, such uncertainties can apply in two classes: uncertainties related to expected background-event rates and to the overall signal-rate expectation.
Since, for the energies of interest, $P(\nu_e \to \nu_s)$ is flat, the goal in a precision measurement is to reduce such uncertainties as much as possible.

\begin{table}[!htbp]
\begin{center}
\caption{Expected signal/background rates under nominal exposures for the PandaX-4T US2 and Paired analyses. We provide expected event rates and our estimate of each associated systematic uncertainty.\label{tab:background_systematics_pandax}}\vspace{0.1em}
\begin{tabular}{|c||c|c||c|c| }\hline
    & \multicolumn{2}{c||}{\textbf{PandaX-4T US2 only}} & \multicolumn{2}{c|}{\textbf{PandaX-4T Paired}} \\ \hline
    & Expectation & Relative Rate & Expectation & Relative Rate  \\ \hline\hline
Signal Rate & $43 \pm 10$ & $23\%$ & $3.4 \pm 1.04$ & $30\%$ \\ \hline
Background 1 & Cathode $204 \pm 45$ & $22\%$ & AC $2.54 \pm 0.73$ & $28\%$ \\ \hline
Background 2 & MD $45\pm 7$ & $16\%$ & Neutron-related $0.1 \pm 0.04$ & $40\%$ \\ \hline
Background 3 & ER $2.2 \pm 0.3$ & $14\%$ & ER + surface $0.17 \pm 0.07$ & $40\%$  \\ \hline \hline
Total & $294.2 \pm 46.6$ & $16\%$ & $6.21 \pm 1.27$ & $20\%$ \\ \hline
\end{tabular}
\end{center}
\end{table}

\begin{table}[!htbp]
\begin{center}
\caption{Expected signal/background rate under nominal exposures for the XENONnT and LZ Paired analyses. As with~\cref{tab:background_systematics_pandax}, we provide expected event rates with associated uncertainties.\label{tab:background_systematics_XENON_LZ}}\vspace{0.1em}
\begin{tabular}{|c||c|c||c|c|} \hline
& \multicolumn{2}{c||}{\textbf{XENONnT}} & \multicolumn{2}{c|}{\textbf{LZ}} \\ \hline
& Expectation & Relative Rate & Expectation & Relative Rate \\ \hline\hline
Signal Rate & $10.7 \pm 3.95$ & $36\%$ & $20.6^{+8.9}_{-6.8}$ & $33-43\%$ \\ \hline
Accidental Coinc. & $25.3 \pm 1.7$ & $6\%$ & $6.6 \pm 0.3$ & $4.5\%$ \\ \hline
Neutron-related & $0.5 \pm 0.3$ & $60\%$ & $0.04^{+0.25}_{-0.04}$ & ${\sim}100\%$ \\ \hline
ER & $0.5 \pm 0.65$ & ${\sim}100\%$ & -- & -- \\ \hline \hline
Total & $37.0 \pm 4.4$ & $12\%$ & $27.2^{+10.1}_{-7.3}$ & $27-37\%$ \\ \hline
\end{tabular}
\end{center}
\end{table}
We summarize the current state of systematic uncertainties on both the expected signal and background rates for the two PandaX-4T analyses in~\cref{tab:background_systematics_pandax} and for XENONnT and LZ's paired analyses in~\cref{tab:background_systematics_XENON_LZ}.
For each class of signal or background, we also provide that systematic uncertainty's relative percentage rate. Finally, we combine these into a total event-rate expectation with a corresponding systematic uncertainty (by adding all uncertainties in quadrature). We see that the experiments presently exhibit $\mathcal{O}(10\%)$ systematic uncertainties in their present analyses -- 16\% overall background uncertainty in PandaX-4T US2 and 12\% in XENONnT Paired. LZ's systematic uncertainty on its total expected rate is dominated by the $\mathcal{O}(30\%)$ uncertainty on the $^{8}$B signal expectation; the uncertainties on its accidental coincidence and neutron-related backgrounds is relatively low.

Reducing each background contribution's relative uncertainty over time will assist in improving the overall systematic uncertainty of such analyses, as will improving the signal-to-background ratios. Ultimately, any future experiment with longer exposures and improved signal-to-background ratios will reach limitations due to systematic uncertainties on its signal expectation. This can be limited by detector effects as well as the theoretical uncertainties behind solar-neutrino model predictions, which presently are at the few-percent level.

In the following section, we will demonstrate how experimental sensitivity is limited by a given systematic-uncertainty floor. In order to be competitive with other sterile-neutrino probes of $|U_{\mu 4}|^2$ and $|U_{\tau 4}|^2$, systematic uncertainties in the $1-5\%$ range will be necessary.

\section{Future CE$\nu$NS Projections}\label{sec:Future}
In this section, we make projections for how future CE$\nu$NS detection of solar neutrinos can constrain sterile-neutrino parameter space.
We take a simplified approach, in light of the above discussion, working in terms of (a) the ultimate systematic uncertainty achievable in an experimental environment as well as (b) the required statistics/exposure necessary to target parameter space of interest.

To motivate this, let us return to Point A, defined in~\cref{eq:PointA}. As we showed in~\cref{fig:Pes_UncertaintyImpact}, this point predicts $P(\nu_e \to \nu_s) \approx 0.05$ for the energy range of interest for $^{8}$B neutrinos.
While this parameter combination (for sufficiently large $\Delta m_{41}^2$) is excluded by various long-baseline neutrino experiments~\cite{Dentler:2018sju}, it is an interesting benchmark against which to compare, as it is \textit{not} currently tested by existing solar-neutrino constraints.

In order to have experimental sensitivity to this point, we must achieve systematic uncertainty below the $5\%$ level and acquire enough statistics so that the relative statistical uncertainty is also below this level.
Whether we have sufficient statistics to reach $5\%$ precision depends both on the exposure of the experiment as well as the signal-to-background ratio achievable. As we saw in the discussion surrounding~\cref{tab:background_systematics_pandax,tab:background_systematics_XENON_LZ}, the various experimental analyses have each already achieved $\mathcal{O}(10-50\%)$ signal-to-background ratios. LZ, in particular, has demonstrated signal-to-background expectations at the rate of approximately $3$-to-$1$. With larger detector volumes and advanced detection techniques, increased signal-to-background ratios are likely.

\begin{figure}[!htbp]
    \centering
\includegraphics[width=\linewidth]{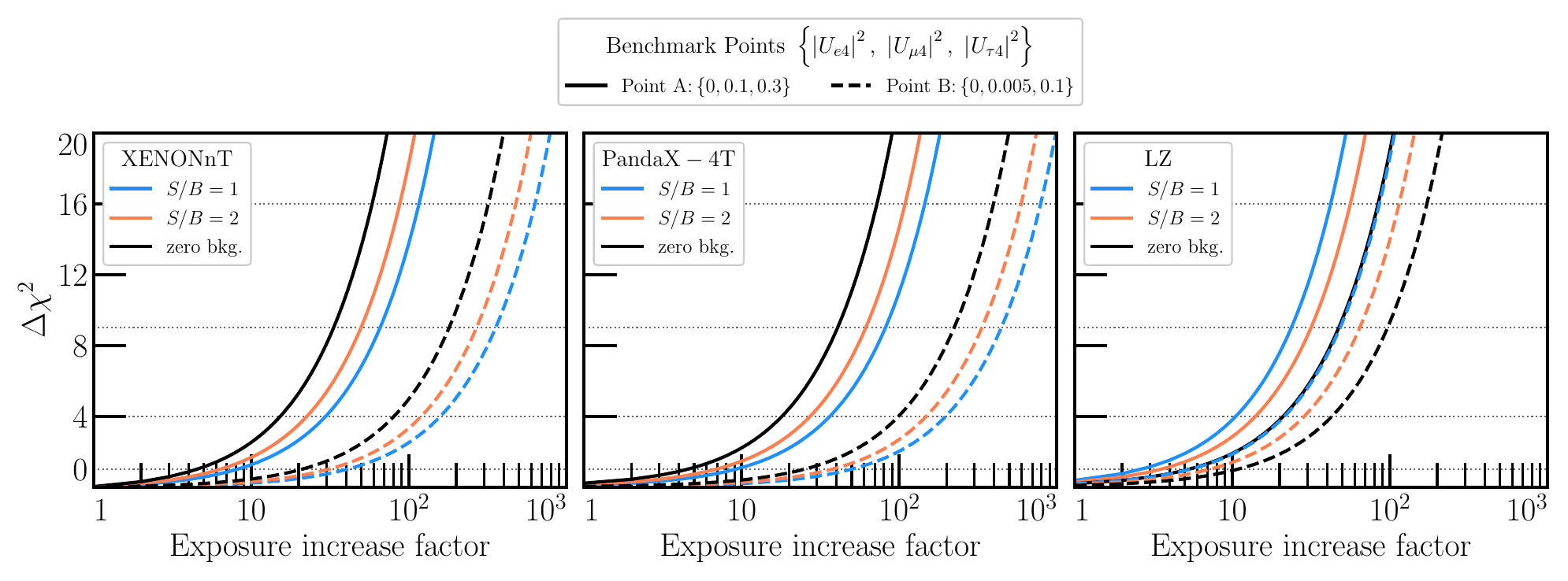}
    \caption{Statistics-only sensitivity to benchmark parameter points `A' and `B' (specified in the text and legend) for solid and dashed lines, respectively, considering an increased-exposure version of three different experiments: XENONnT (left), PandaX-4T (center), and LZ (right). We show projections for increased exposures for a variety of signal-to-background ratios: a perfect, zero-background measurement (black), and $S/B = 1$ (blue) or $2$ (orange). In general, these experiments can reach these benchmark points (statistically) for $\mathcal{O}(10-100)$ factors of increase in exposure.}
    \label{fig:ExposureScaling_XENON}
\end{figure}

\textbf{Statistical Uncertainty Benchmarks:} Nominally, a $5\%$ statistical uncertainty is achieved with $\mathcal{O}(600)$ signal events, a factor of $10-50$ larger exposure than what the three experiments have collected to date.
In~\cref{fig:ExposureScaling_XENON}, we demonstrate how an increase in exposure translates into the required \textit{statistics-only} uncertainty to test Point A (solid lines) and Point B (dashed) relative to what each of the three analyses has observed to date -- the left panel corresponds to XENONnT, the middle PandaX-4T, and the right LZ.
In each case, we allow for the signal-to-background ratio to vary as well -- zero-background estimates are shown in black, and signal-to-background ratios of $1$ (equal signal and background) and $2$ (two signal events for each background) are shown in blue and orange, respectively.
For the more-reasonably-accessible Point A, we see that, regardless of the signal-to-background ratio, these experiments can achieve sufficient statistics in factors of ${\sim}10-30$ increases of exposure - this corresponds to $\mathcal{O}(50)$ ton-yrs of exposure, certainly within the realm of possibility with current and next-generation facilities.

Accessing the more exciting Point B (with substantially smaller sterile-neutrino mixing angles) will be a more substantial challenge. However, this seems possible with approximately a factor of $100$ more exposure, or approximately several hundred ton-yrs of exposure total. This type of exposure could be possible in future experiments. We note that, while a signal-to-background ratio of $2$ is currently high from the perspective of XENONnT and PandaX-4T's US2 analyses, LZ has already achieved a signal-to-background ratio of approximately $3$. Improvements to background-rejection techniques, as well as larger detector volumes (which can reduce neutron-related signatures relative to signal rates), should allow such ratios to continue improving with time.
\vspace{0.5em}

\begin{figure}[!htbp]
    \centering
\includegraphics[width=0.7\linewidth]{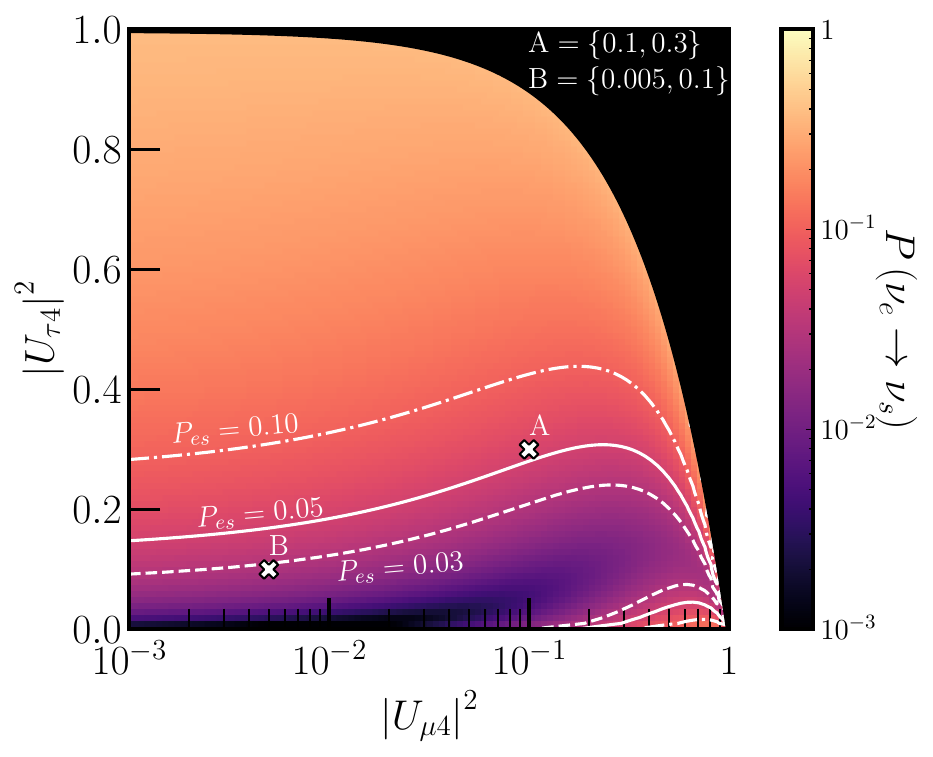}
    \caption{The oscillation probability for solar $\nu_e$ into sterile neutrinos $\nu_s$ as a function of $\left\lvert U_{\mu 4}\right\rvert^2$ vs.~$\left\lvert U_{\tau 4}\right\rvert^2$ (with $\left\lvert U_{e4}\right\rvert^2 = 0$) -- black regions corresponding to smaller probabilities (i.e.~closer to the standard three-flavor oscillatino scenario), and orange corresponding to larger ones. Isocontours of $P(\nu_e \to \nu_s)$ are indicated as white lines for $3\%$ (dashed), $5\%$ (solid), and $10\%$ (dot-dashed) probabilities -- benchmark parameter points `A' and `B,' discussed throughout the text, are shown for comparison. The oscillation probability sets a systematic-uncertainty goal for a future experiment.\label{fig:probability Pes heatmap}}
\end{figure}

\textbf{Systematic Uncertainty Benchmarks:} 
Previously, we demonstrated that Point A and Point B, two interesting points in parameter space, corresponded to oscillation probabilities $P(\nu_e \to \nu_s)$ of approximately $5\%$ and $3\%$, respectively. Achieving sensitivity to these points in parameter space required sufficient statistics (including low enough background rates) for detection, as well as reduction of systematic uncertainties to at least this level. In~\cref{fig:probability Pes heatmap}, we show the oscillation probability $P(\nu_e \to \nu_s)$ for a wider range of $\left\lvert U_{\mu 4}\right\rvert^2$ and $\left\lvert U_{\tau 4}\right\rvert^2$, with Points A and B labeled accordingly.
The various colors indicate different values of $P(\nu_e \to \nu_s)$, and different white lines indicate iso-contours of $P(\nu_e \to \nu_s) = 0.1$ (dot-dashed), $0.05$ (solid), and $0.03$ (dashed). We can therefore expect that, given a particular experimental exposure (as well as assumptions regarding signal and background rates) and a particular level of systematic uncertainty, that experimental sensitivity will lie along one such contour in this space. For instance, if $3\%$ systematic uncertainties are attained (and this is the limiting factor; statistical uncertainties are smaller than this), then we expect an experiment to achieve sensitivity at Point B and everywhere along the dashed white line.

\begin{figure}[!htbp]
    \centering
\includegraphics[width=0.7\linewidth]{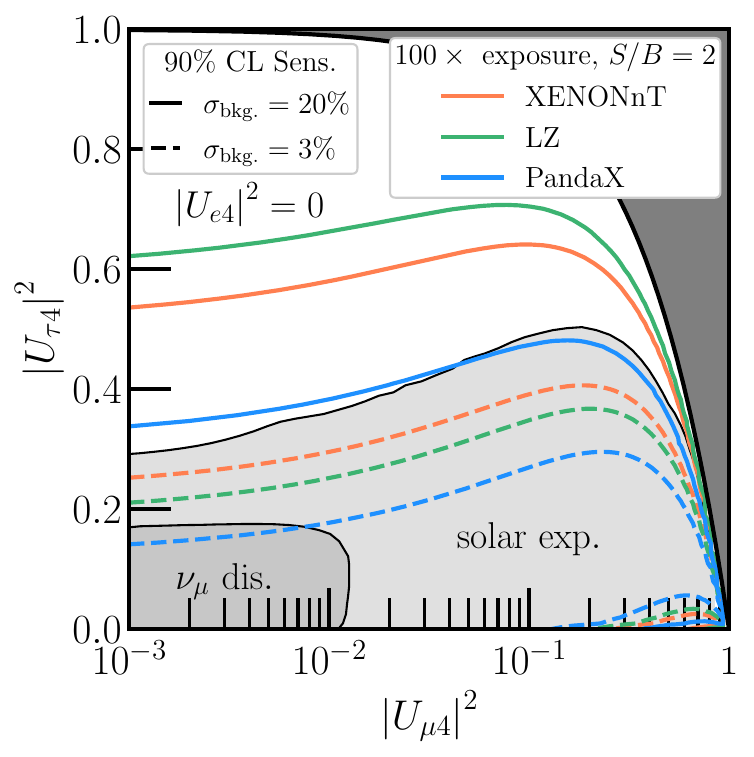}
    \caption{Potential sensitivity to $\left\lvert U_{\mu 4}\right\rvert^2$ vs.~$\left\lvert U_{\tau 4}\right\rvert^2$ (with $\left\lvert U_{e4}\right\rvert^2 = 0$) for increased exposures of XENONnT (orange), LZ (green), and PandaX (blue), assuming a factor of 100 larger exposure. All projections assume a signal-to-background ratio of $2$, and solid (dashed) lines correspond to background systematics constrained to the $20\%$ ($3\%$) level. The gray, labelled regions correspond to constraints from solar experiments only or from all muon-neutrino disappearance, from Ref.~\cite{Dentler:2018sju}. \label{fig:100x sig-bkg=2 x and p}}
\end{figure}

We perform a projection for future sensitivity by enhanced exposures of each of XENONnT, PandaX-4T, and LZ, each rescaled by a factor of $100$, and derive 90\% CL sensitivity under two assumptions: one in which the background systematic uncertainty is controlled at the 20\% level, and one in which it can be decreased to the 3\% level. The results of this are shown in~\cref{fig:100x sig-bkg=2 x and p} for XENONnT (orange), PandaX-4T (blue) and LZ (green), as a function of $\left\lvert U_{\mu 4}\right\rvert^2$ and $\left\lvert U_{\tau 4}\right\rvert^2$, all with $\left\lvert U_{e4}\right\rvert^2$ fixed to zero.
We assume that there is negligible signal normalization uncertainty present, and that a signal-to-background ratio of $2$ is achievable for all analyses.
Unsurprisingly, our future PandaX projection offers the strongest sensitivity; this is expected due to the current high $^{8}$B-rate expectation (43 events) relative to the expectations from XENONnT (10.7) and LZ (20.6). This is partially because we have projected forward PandaX's US2-only analysis. If instead, we projected forward its S1/S2 paired analysis, it would be weaker than both LZ and XENONnT.
We also directly compare against existing experimental constraints, singling out specifically constraints that come from solar neutrino experiments (light gray) -- driven largely by Super-Kamiokande~\cite{Super-Kamiokande:2005wtt,Super-Kamiokande:2008ecj,Super-Kamiokande:2010tar}, SNO~\cite{SNO:2005oxr,SNO:2006odc,SNO:2008gqy}, and Borexino~\cite{Bellini:2011rx,Borexino:2008fkj} -- as well as a global combination of $\nu_\mu$/$\bar\nu_\mu$ disappearance experiments~\cite{Dentler:2018sju}.

Concretely, we see that with $3\%$ background uncertainties, $\mathcal{O}(500)$ ton-yr exposures, and signal-to-background ratios of $2$ or higher, direct detection facilities may exploit CE$\nu$NS to explore new sterile-neutrino parameter space, even space not accessed by high-intensity $\nu_\mu$-beam-based measurements. 

\section{Discussion \& Conclusions}\label{sec:Conclusions}
Solar ${}^8$B CE$\nu$NS has now been detected by three dark matter experiments: PandaX-4T, XENONnT, and LZ. This is an exciting detection that represents the first identification of CE$\nu$NS from an astrophysical source, opening a new avenue for study of neutrino physics in dark-matter detectors. Each of these measurements are broadly consistent with one another, with uncertainties on the measured fluxes that are comparable across the experiments. Current analyses are limited by ${\cal O(}10\%)$ total-rate systematics in PandaX-4T (US2 16\%, Paired $\sim$ 20\%) and XENONnT (12\%), and by a larger uncertainty on the $^{8}$B signal normalization in LZ (contributing to a $\sim$ 27–37\% total).

We have examined how these measurements and future such measurements can probe mixing of a light sterile state with the muon and tau flavors. In this scenario we have shown that the primary observable for $^8$B solar neutrinos is an overall rate normalization governed by $1-P(\nu_e \to \nu_s)$. There are two important motivations for the sterile analysis. First, because $P(\nu_e \rightarrow \nu_e$) cannot probe $|U_{\mu 4}|^2$ or $|U_{\tau 4}|^2$ when $|U_{e4}|^2 \to 0$, CE$\nu$NS offers a unique neutral-current handle on active-to-sterile conversion that is directly sensitive to muon and tau admixtures through $P(\nu_e \rightarrow \nu_s)$. Second, the interpretation of any departure from the standard expectation is impacted by uncertainties in the standard oscillation parameters, especially $\theta_{23}$ and $\delta_{\rm CP}$; improved external constraints on these parameters, for example from DUNE and/or T2HK, will shrink these predicted bands and improve inferences from solar CE$\nu$NS-based experiments.  

Since the main ${}^8$B-driven observable is a rate normalization, sensitivity is limited by systematic uncertainties. We have examined these systematics in detail and find that reaching competitiveness with existing sterile searches requires total systematics at the few-percent level and exposures of $\mathcal{O}(10^2-10^3)$ ton-years with signal-to-background ratio $\sim 1-2$. Assuming background systematics at the $\sim 3\%$ level and signal-to-background $\gtrsim 2$, we find that exposures of $\sim 500$ ton-years can begin to probe sterile-mixing parameter space not accessible to current solar or even $\nu_\mu/ \bar{\nu_\mu}$ disappearance data. Next-generation programs approaching $\sim 3000$ ton-years could provide complementary coverage of $|U_{\mu4}|^2$-$|U_{\tau4}|^2$ space. Achieving this potential will require improvements in background modeling and rejection, calibration of detector response and signal normalization, and the incorporation of updated three-flavor constraints to limit theoretical uncertainties in $P(\nu_e \rightarrow \nu_s)$. 

More generally, our analysis emphasizes that solar CE$\nu$NS provides a uniquely clean window on the low-energy muon and tau components of the ${}^8$B flux. At the relevant MeV energies, charged-current production of muons and taus is kinematically forbidden, so conventional solar measurements predominantly access $\nu_e$ (via CC) or are only weakly sensitive to $\nu_{\mu,\tau}$ (via ES with reduced cross sections), whereas CE$\nu$NS is a neutral-current process that is nearly flavor blind and coherently enhanced, rendering the measured rate directly proportional to the total active flux. Future experiments as we have discussed are the only viable methods to study differences between the mu and tau neutrino fluxes at these energies, and are complementary to more traditional measurements of these fluxes from higher energy sources. 

While the focus of our analysis has been on identification of sterile neutrinos with CE$\nu$NS, in the future it will also be possible to combine with electron scattering data to search for sterile neutrinos from the Sun through the neutrino-electron elastic scattering channel. Indeed, previous projections have shown strong sensitivity to $|U_{e4}|^2$ via this channel, driven by the substantial event rate~\citep{Harnik:2012ni,Billard:2013qya}. Since this involves electron detection backgrounds will be different from those considered in this paper, though ideally such a signal could be combined with the CE$\nu$NS channel for optimal sensitivity.   

\section*{Acknowledgments} 
The work of KJK, NM, and LES is supported in part by US DOE Grant \#DE-SC0010813. We thank Pedro Machado for valuable conversations regarding this work.

\bibliographystyle{utphys}
\bibliography{main}

@article{COHERENT:2024axu,
    author = "Adamski, S. and others",
    collaboration = "COHERENT",
    title = "{First detection of coherent elastic neutrino-nucleus scattering on germanium}",
    eprint = "2406.13806",
    archivePrefix = "arXiv",
    primaryClass = "hep-ex",
    month = "6",
    year = "2024"
}

@article{COHERENT:2021xmm,
    author = "Akimov, D. and others",
    collaboration = "COHERENT",
    title = "{Measurement of the Coherent Elastic Neutrino-Nucleus Scattering Cross Section on CsI by COHERENT}",
    eprint = "2110.07730",
    archivePrefix = "arXiv",
    primaryClass = "hep-ex",
    doi = "10.1103/PhysRevLett.129.081801",
    journal = "Phys. Rev. Lett.",
    volume = "129",
    number = "8",
    pages = "081801",
    year = "2022"
}

@article{Dentler:2018sju,
    author = "Dentler, Mona and Hern{\'a}ndez-Cabezudo, {\'A}lvaro and Kopp, Joachim and Machado, Pedro A. N. and Maltoni, Michele and Martinez-Soler, Ivan and Schwetz, Thomas",
    title = "{Updated Global Analysis of Neutrino Oscillations in the Presence of eV-Scale Sterile Neutrinos}",
    eprint = "1803.10661",
    archivePrefix = "arXiv",
    primaryClass = "hep-ph",
    reportNumber = "MITP-18-023, IFT-UAM-CSIC-18-033, MITP/18-023, FERMILAB-PUB-18-086-T, IFT-UAM/CSIC-18-033",
    doi = "10.1007/JHEP08(2018)010",
    journal = "JHEP",
    volume = "08",
    pages = "010",
    year = "2018"
}

@article{COHERENT:2020iec,
    author = "Akimov, D. and others",
    collaboration = "COHERENT",
    title = "{First Measurement of Coherent Elastic Neutrino-Nucleus Scattering on Argon}",
    eprint = "2003.10630",
    archivePrefix = "arXiv",
    primaryClass = "nucl-ex",
    doi = "10.1103/PhysRevLett.126.012002",
    journal = "Phys. Rev. Lett.",
    volume = "126",
    number = "1",
    pages = "012002",
    year = "2021"
}

@article{XENON:2024ijk,
    author = "Aprile, Elena and others",
    collaboration = "XENON",
    title = "{First Indication of Solar B8 Neutrinos via Coherent Elastic Neutrino-Nucleus Scattering with XENONnT}",
    eprint = "2408.02877",
    archivePrefix = "arXiv",
    primaryClass = "nucl-ex",
    doi = "10.1103/PhysRevLett.133.191002",
    journal = "Phys. Rev. Lett.",
    volume = "133",
    number = "19",
    pages = "191002",
    year = "2024"
}

@article{PandaX:2024muv,
    author = "Bo, Zihao and others",
    collaboration = "PandaX",
    title = "{First Indication of Solar B8 Neutrinos through Coherent Elastic Neutrino-Nucleus Scattering in PandaX-4T}",
    eprint = "2407.10892",
    archivePrefix = "arXiv",
    primaryClass = "hep-ex",
    doi = "10.1103/PhysRevLett.133.191001",
    journal = "Phys. Rev. Lett.",
    volume = "133",
    number = "19",
    pages = "191001",
    year = "2024"
}

@article{Ackermann:2025obx,
    author = "Ackermann, N. and others",
    title = "{First observation of reactor antineutrinos by coherent scattering}",
    eprint = "2501.05206",
    archivePrefix = "arXiv",
    primaryClass = "hep-ex",
    month = "1",
    year = "2025"
}

@article{Abdullah:2022zue,
    author = "Abdullah, M. and others",
    title = "{Coherent elastic neutrino-nucleus scattering: Terrestrial and astrophysical applications}",
    eprint = "2203.07361",
    archivePrefix = "arXiv",
    primaryClass = "hep-ph",
    month = "3",
    year = "2022"
}

@article{Duque:2025mno,
    author = "Duque, Laura and Lamprea, J. M. and Miranda, Omar G.",
    title = "{Shifting the neutrino fog: studying the Isospin-violating Dark Matter case}",
    eprint = "2512.19784",
    archivePrefix = "arXiv",
    primaryClass = "hep-ph",
    month = "12",
    year = "2025"
}

@article{DeRomeri:2024iaw,
    author = "De Romeri, Valentina and Papoulias, Dimitrios K. and Ternes, Christoph A.",
    title = "{Bounds on new neutrino interactions from the first CE{\ensuremath{\nu}}NS data at direct detection experiments}",
    eprint = "2411.11749",
    archivePrefix = "arXiv",
    primaryClass = "hep-ph",
    doi = "10.1088/1475-7516/2025/05/012",
    journal = "JCAP",
    volume = "05",
    pages = "012",
    year = "2025"
}

@article{XENON:2024xgd,
    author = "Aprile, E. and others",
    collaboration = "XENON, (XENON Collaboration){\textdaggerdbl}{\textdaggerdbl}",
    title = "{XENONnT WIMP search: Signal and background modeling and statistical inference}",
    eprint = "2406.13638",
    archivePrefix = "arXiv",
    primaryClass = "physics.data-an",
    doi = "10.1103/PhysRevD.111.103040",
    journal = "Phys. Rev. D",
    volume = "111",
    number = "10",
    pages = "103040",
    year = "2025"
}

@article{AristizabalSierra:2024nwf,
    author = "Aristizabal Sierra, D. and Mishra, N. and Strigari, L.",
    title = "{Implications of first neutrino-induced nuclear recoil measurements in direct detection experiments: Probing nonstandard interaction via CE{\ensuremath{\nu}}NS}",
    eprint = "2409.02003",
    archivePrefix = "arXiv",
    primaryClass = "hep-ph",
    doi = "10.1103/PhysRevD.111.055007",
    journal = "Phys. Rev. D",
    volume = "111",
    number = "5",
    pages = "055007",
    year = "2025"
}

@article{DeRomeri:2024hvc,
    author = "De Romeri, Valentina and Papoulias, Dimitrios K. and Sanchez Garcia, Gonzalo and Ternes, Christoph A. and T{\'o}rtola, Mariam",
    title = "{Neutrino electromagnetic properties and sterile dipole portal in light of the first solar CE{\ensuremath{\nu}}NS~data}",
    eprint = "2412.14991",
    archivePrefix = "arXiv",
    primaryClass = "hep-ph",
    doi = "10.1088/1475-7516/2025/05/080",
    journal = "JCAP",
    volume = "05",
    pages = "080",
    year = "2025"
}

@article{Blanco-Mas:2024ale,
    author = "Blanco-Mas, Pablo and Coloma, Pilar and Herrera, Gonzalo and Huber, Patrick and Kopp, Joachim and Shoemaker, Ian M. and Tabrizi, Zahra",
    title = "{Clarity through the Neutrino Fog: Constraining New Forces in Dark Matter Detectors}",
    eprint = "2411.14206",
    archivePrefix = "arXiv",
    primaryClass = "hep-ph",
    reportNumber = "IFT-UAM/CSIC-24-164",
    month = "11",
    year = "2024"
}

@article{Li:2024iij,
    author = "Li, Gang and Song, Chuan-Qiang and Tang, Feng-Jie and Yu, Jiang-Hao",
    title = "{Constraints on neutrino nonstandard interactions from COHERENT, PandaX-4T and XENONnT}",
    eprint = "2409.04703",
    archivePrefix = "arXiv",
    primaryClass = "hep-ph",
    doi = "10.1103/PhysRevD.111.035002",
    journal = "Phys. Rev. D",
    volume = "111",
    number = "3",
    pages = "035002",
    year = "2025"
}

@article{Tomalak:2020zfh,
    author = "Tomalak, Oleksandr and Machado, Pedro and Pandey, Vishvas and Plestid, Ryan",
    title = "{Flavor-dependent radiative corrections in coherent elastic neutrino-nucleus scattering}",
    eprint = "2011.05960",
    archivePrefix = "arXiv",
    primaryClass = "hep-ph",
    reportNumber = "FERMILAB-PUB-20-524-T",
    doi = "10.1007/JHEP02(2021)097",
    journal = "JHEP",
    volume = "02",
    pages = "097",
    year = "2021"
}

@article{OHare:2021utq,
    author = "O'Hare, Ciaran A. J.",
    title = "{New Definition of the Neutrino Floor for Direct Dark Matter Searches}",
    eprint = "2109.03116",
    archivePrefix = "arXiv",
    primaryClass = "hep-ph",
    doi = "10.1103/PhysRevLett.127.251802",
    journal = "Phys. Rev. Lett.",
    volume = "127",
    number = "25",
    pages = "251802",
    year = "2021"
}

@article{Esteban:2024eli,
    author = "Esteban, Ivan and Gonzalez-Garcia, M. C. and Maltoni, Michele and Martinez-Soler, Ivan and Pinheiro, Jo{\~a}o Paulo and Schwetz, Thomas",
    title = "{NuFit-6.0: updated global analysis of three-flavor neutrino oscillations}",
    eprint = "2410.05380",
    archivePrefix = "arXiv",
    primaryClass = "hep-ph",
    reportNumber = "IFT-UAM/CSIC-24-140, YITP-SB-2024-24, IPPP/24/64, IPPP/24/64, IFT-UAM/CSIC-24-140, YITP-SB-2024-24",
    doi = "10.1007/JHEP12(2024)216",
    journal = "JHEP",
    volume = "12",
    pages = "216",
    year = "2024"
}

@article{Billard:2014yka,
    author = "Billard, J. and Strigari, L. E. and Figueroa-Feliciano, E.",
    title = "{Solar neutrino physics with low-threshold dark matter detectors}",
    eprint = "1409.0050",
    archivePrefix = "arXiv",
    primaryClass = "astro-ph.CO",
    doi = "10.1103/PhysRevD.91.095023",
    journal = "Phys. Rev. D",
    volume = "91",
    number = "9",
    pages = "095023",
    year = "2015"
}

@article{Dutta:2019oaj,
    author = "Dutta, Bhaskar and Strigari, Louis E.",
    title = "{Neutrino physics with dark matter detectors}",
    eprint = "1901.08876",
    archivePrefix = "arXiv",
    primaryClass = "hep-ph",
    doi = "10.1146/annurev-nucl-101918-023450",
    journal = "Ann. Rev. Nucl. Part. Sci.",
    volume = "69",
    pages = "137--161",
    year = "2019"
}

@article{Billard:2013qya,
    author = "Billard, J. and Strigari, L. and Figueroa-Feliciano, E.",
    title = "{Implication of neutrino backgrounds on the reach of next generation dark matter direct detection experiments}",
    eprint = "1307.5458",
    archivePrefix = "arXiv",
    primaryClass = "hep-ph",
    doi = "10.1103/PhysRevD.89.023524",
    journal = "Phys. Rev. D",
    volume = "89",
    number = "2",
    pages = "023524",
    year = "2014"
}

@article{Ruppin:2014bra,
    author = "Ruppin, F. and Billard, J. and Figueroa-Feliciano, E. and Strigari, L.",
    title = "{Complementarity of dark matter detectors in light of the neutrino background}",
    eprint = "1408.3581",
    archivePrefix = "arXiv",
    primaryClass = "hep-ph",
    doi = "10.1103/PhysRevD.90.083510",
    journal = "Phys. Rev. D",
    volume = "90",
    number = "8",
    pages = "083510",
    year = "2014"
}

@article{Sehgal:1985iu,
    author = "Sehgal, L. M.",
    title = "{Differences in the Coherent Interactions of $\nu_e$, $\nu_\mu$ and $\nu_\tau$}",
    reportNumber = "PITHA-85-13-REV, PITHA-85-13",
    doi = "10.1016/0370-2693(85)90942-6",
    journal = "Phys. Lett. B",
    volume = "162",
    pages = "370--372",
    year = "1985"
}

@article{XENON:2024hup,
    author = "Aprile, E. and others",
    collaboration = "XENON",
    title = "{First Search for Light Dark Matter in the Neutrino Fog with XENONnT}",
    eprint = "2409.17868",
    archivePrefix = "arXiv",
    primaryClass = "hep-ex",
    doi = "10.1103/PhysRevLett.134.111802",
    journal = "Phys. Rev. Lett.",
    volume = "134",
    number = "11",
    pages = "111802",
    year = "2025"
}

@article{LZ:2024zvo,
    author = "Aalbers, J. and others",
    collaboration = "LZ",
    title = "{Dark Matter Search Results from 4.2{\,}{\,}Tonne-Years of Exposure of the LUX-ZEPLIN (LZ) Experiment}",
    eprint = "2410.17036",
    archivePrefix = "arXiv",
    primaryClass = "hep-ex",
    reportNumber = "FERMILAB-PUB-24-0796-V",
    doi = "10.1103/4dyc-z8zf",
    journal = "Phys. Rev. Lett.",
    volume = "135",
    number = "1",
    pages = "011802",
    year = "2025"
}

@article{Gonzalez-Garcia:2024hmf,
    author = "Gonzalez-Garcia, M. C. and Maltoni, Michele and Pinheiro, Jo{\~a}o Paulo",
    title = "{Solar model independent constraints on the sterile neutrino interpretation of the Gallium Anomaly}",
    eprint = "2411.16840",
    archivePrefix = "arXiv",
    primaryClass = "hep-ph",
    reportNumber = "YITP-SB-2024-29, IFT-UAM/CSIC-24-166",
    doi = "10.1016/j.physletb.2025.139297",
    journal = "Phys. Lett. B",
    volume = "862",
    pages = "139297",
    year = "2025"
}

@article{LZ:2025igz,
    author = "Akerib, D. S. and others",
    collaboration = "LZ",
    title = "{Searches for Light Dark Matter and Evidence of Coherent Elastic Neutrino-Nucleus Scattering of Solar Neutrinos with the LUX-ZEPLIN (LZ) Experiment}",
    eprint = "2512.08065",
    archivePrefix = "arXiv",
    primaryClass = "hep-ex",
    month = "12",
    year = "2025"
}

@article{XENON:2026ydt,
    author = "Aprile, E. and others",
    collaboration = "XENON",
    title = "{Probing the Solar $^8$B Neutrino Fog with XENONnT}",
    eprint = "2604.06002",
    archivePrefix = "arXiv",
    primaryClass = "hep-ex",
    month = "4",
    year = "2026"
}

@article{PandaX:2024qfu,
    author = "Bo, Zihao and others",
    collaboration = "PandaX",
    title = "{Dark Matter Search Results from 1.54{\,}{\,}Tonne{\textperiodcentered}Year Exposure of PandaX-4T}",
    eprint = "2408.00664",
    archivePrefix = "arXiv",
    primaryClass = "hep-ex",
    doi = "10.1103/PhysRevLett.134.011805",
    journal = "Phys. Rev. Lett.",
    volume = "134",
    number = "1",
    pages = "011805",
    year = "2025"
}

@article{DeRomeri:2026prc,
    author = "De Romeri, Valentina and Papoulias, Dimitrios K. and Pompa, Federica and Sanchez Garcia, Gonzalo and Ternes, Christoph A.",
    title = "{Testing light and heavy vector mediators with solar CE$\nu$NS measurements}",
    eprint = "2603.00554",
    archivePrefix = "arXiv",
    primaryClass = "hep-ph",
    month = "2",
    year = "2026"
}

@article{Super-Kamiokande:2005wtt,
    author = "Hosaka, J. and others",
    collaboration = "Super-Kamiokande",
    title = "{Solar neutrino measurements in super-Kamiokande-I}",
    eprint = "hep-ex/0508053",
    archivePrefix = "arXiv",
    doi = "10.1103/PhysRevD.73.112001",
    journal = "Phys. Rev. D",
    volume = "73",
    pages = "112001",
    year = "2006"
}

@article{Super-Kamiokande:2008ecj,
    author = "Cravens, J. P. and others",
    collaboration = "Super-Kamiokande",
    title = "{Solar neutrino measurements in Super-Kamiokande-II}",
    eprint = "0803.4312",
    archivePrefix = "arXiv",
    primaryClass = "hep-ex",
    doi = "10.1103/PhysRevD.78.032002",
    journal = "Phys. Rev. D",
    volume = "78",
    pages = "032002",
    year = "2008"
}

@article{Super-Kamiokande:2010tar,
    author = "Abe, K. and others",
    collaboration = "Super-Kamiokande",
    title = "{Solar neutrino results in Super-Kamiokande-III}",
    eprint = "1010.0118",
    archivePrefix = "arXiv",
    primaryClass = "hep-ex",
    doi = "10.1103/PhysRevD.83.052010",
    journal = "Phys. Rev. D",
    volume = "83",
    pages = "052010",
    year = "2011"
}

@article{SNO:2006odc,
    author = "Aharmim, B. and others",
    collaboration = "SNO",
    title = "{Determination of the $\nu_e$ and total $^8$B solar neutrino fluxes with the Sudbury neutrino observatory phase I data set}",
    eprint = "nucl-ex/0610020",
    archivePrefix = "arXiv",
    doi = "10.1103/PhysRevC.75.045502",
    journal = "Phys. Rev. C",
    volume = "75",
    pages = "045502",
    year = "2007"
}

@article{SNO:2005oxr,
    author = "Aharmim, B. and others",
    collaboration = "SNO",
    title = "{Electron energy spectra, fluxes, and day-night asymmetries of B-8 solar neutrinos from measurements with NaCl dissolved in the heavy-water detector at the Sudbury Neutrino Observatory}",
    eprint = "nucl-ex/0502021",
    archivePrefix = "arXiv",
    doi = "10.1103/PhysRevC.72.055502",
    journal = "Phys. Rev. C",
    volume = "72",
    pages = "055502",
    year = "2005"
}

@article{SNO:2008gqy,
    author = "Aharmim, B. and others",
    collaboration = "SNO",
    title = "{An Independent Measurement of the Total Active B-8 Solar Neutrino Flux Using an Array of He-3 Proportional Counters at the Sudbury Neutrino Observatory}",
    eprint = "0806.0989",
    archivePrefix = "arXiv",
    primaryClass = "nucl-ex",
    reportNumber = "LA-UR-08-1316",
    doi = "10.1103/PhysRevLett.101.111301",
    journal = "Phys. Rev. Lett.",
    volume = "101",
    pages = "111301",
    year = "2008"
}

@article{Bellini:2011rx,
    author = "Bellini, G. and others",
    title = "{Precision measurement of the 7Be solar neutrino interaction rate in Borexino}",
    eprint = "1104.1816",
    archivePrefix = "arXiv",
    primaryClass = "hep-ex",
    doi = "10.1103/PhysRevLett.107.141302",
    journal = "Phys. Rev. Lett.",
    volume = "107",
    pages = "141302",
    year = "2011"
}

@article{Borexino:2008fkj,
    author = "Bellini, G. and others",
    collaboration = "Borexino",
    title = "{Measurement of the solar 8B neutrino rate with a liquid scintillator target and 3 MeV energy threshold in the Borexino detector}",
    eprint = "0808.2868",
    archivePrefix = "arXiv",
    primaryClass = "astro-ph",
    doi = "10.1103/PhysRevD.82.033006",
    journal = "Phys. Rev. D",
    volume = "82",
    pages = "033006",
    year = "2010"
}

@article{Harnik:2012ni,
    author = "Harnik, Roni and Kopp, Joachim and Machado, Pedro A. N.",
    title = "{Exploring nu Signals in Dark Matter Detectors}",
    eprint = "1202.6073",
    archivePrefix = "arXiv",
    primaryClass = "hep-ph",
    reportNumber = "FERMILAB-PUB-12-048-T",
    doi = "10.1088/1475-7516/2012/07/026",
    journal = "JCAP",
    volume = "07",
    pages = "026",
    year = "2012"
}

@article{Mishra:2023jlq,
    author = "Mishra, Nityasa and Strigari, Louis E.",
    title = "{Solar neutrinos with CE{\ensuremath{\nu}}NS and flavor-dependent radiative corrections}",
    eprint = "2305.17827",
    archivePrefix = "arXiv",
    primaryClass = "hep-ph",
    reportNumber = "MI-HET-812",
    doi = "10.1103/PhysRevD.108.063023",
    journal = "Phys. Rev. D",
    volume = "108",
    number = "6",
    pages = "063023",
    year = "2023"
}

@article{Goldhagen_2022,
   title={Testing sterile neutrino mixing with present and future solar neutrino data},
   volume={82},
   ISSN={1434-6052},
   url={http://dx.doi.org/10.1140/epjc/s10052-022-10052-2},
   DOI={10.1140/epjc/s10052-022-10052-2},
   number={2},
   journal={The European Physical Journal C},
   publisher={Springer Science and Business Media LLC},
   author={Goldhagen, Kim and Maltoni, Michele and Reichard, Shayne E. and Schwetz, Thomas},
   year={2022}
}
\end{document}